\documentclass[aps,prb,superscriptaddress,twocolumn,showkeys,amsmath,amssymb,floatfix]{revtex4-2}

\usepackage{amsmath,amssymb,bbm,mathrsfs,bm,braket,color,graphicx,comment,amsfonts,dsfont}
\usepackage{graphicx} % Required for inserting images
\usepackage[colorlinks,citecolor=blue,urlcolor=blue]{hyperref}

\date{\today}

\begin{document}
%\title{Temperature-variation of the zero bias conductance as a probe for structural phase transitions}
\title{Angularly Selective Enhanced Vortex Screening in Extremely  Layered Superconductors with Tilted Columnar Defects}

\author{Gonzalo Rumi}%
\affiliation{Low Temperatures Lab, Centro At\'{o}mico Bariloche, CNEA, Argentina.}
\affiliation{Instituto de Nanociencia y Nanotecnología, CONICET-CNEA, Nodo Bariloche, Argentina.}
\affiliation{Instituto Balseiro,
CNEA and Universidad Nacional de Cuyo, 
Bariloche, Argentina}
\affiliation{Neutrons Beam National Lab,
CNEA, Argentina}

\author{Vincent Mosser}
\affiliation{Vmnetic SAS, F-75008 Paris, France}

\author{Marcin Konczykowski}
\affiliation{Laboratoire des Solides Irradi\'es, CEA/DRF/IRAMIS,
\'{E}cole Polytechnique, CNRS, \mbox{Institut Polytechnique de Paris}, F-91128 Palaiseau, France}

\author{Yanina Fasano}%
\affiliation{Low Temperatures Lab, Centro At\'{o}mico Bariloche, CNEA, Argentina.}
\affiliation{Instituto de Nanociencia y Nanotecnología, CONICET-CNEA, Nodo Bariloche, Argentina.}
\affiliation{Instituto Balseiro,
CNEA and Universidad Nacional de Cuyo, 
Bariloche, Argentina}

\date{\today}

\begin{abstract}
We report on two mechanisms of angularly selective enhanced screening in the solid
vortex phase of extremely layered superconductors
with tilted columnar defects (CDs). We study Bi$_2$Sr$_2$CaCu$_2$O$_{8+\delta}$ samples with different densities of CD tilted 45$^{\circ}$ from the $c$-axis, and conduct local ac Hall magnetometry measurements, probing the sustainable current of the vortex system.
We reveal two types of maxima in sustainable current for particular directions, detected as dips in the magnetic transmittivity of the vortex system. First, for a smaller number of vortices than of defects, an enhancement of screening  is detected at an angular location  $\Theta^{1}_{\rm dip}$$\sim$45$^{\circ}$ for $H$ applied close to the
direction of CD.  For a larger number of vortices than of CD, $\Theta^{1}_{\rm dip}$ decreases towards the $ab$-plane direction upon warming. Second, a pair of additional dips in transmittivity are detected at angles $\Theta^{2}_{dip}$ closer to, and quite symmetric with, the $ab$-plane. These two types of angularly selective enhanced screening  reveal the effective pinning by tilted CD even for the composite vortex lattices nucleated in tilted fields in Bi$_2$Sr$_2$CaCu$_2$O$_{8+\delta}$.
\end{abstract}

%\pacs{$74.25.Uv,74.25.Ha,74.25.Dw$} \keywords{}
 \maketitle

\section{Introduction}

In extremely layered superconductors, when the magnetic
field is applied  tilted with respect to the $c-$axis, the vortex
lattice results in a compound lattice comprising pancake and Josephson
vortices~\cite{Blatter,Bending2005, Koshelev_2005}. Pancake vortices are confined to the CuO$_2$ planes~\cite{Clemm1991, Bulaevskii_1992} and contain a core within that layer where the superconductivity is depleted. On the other hand, Josephson
vortices lack this core structure and are generated by a field component in the \textit{ab-}plane~\cite{Koshelev1999,Koshelev2006,Composite_PV-JV}
The in-plane
lattice spacing of the pancake vortices $a$ is determined by the
$c$-axis field component and that of the Josephson vortices  is
given by the in-plane field component.  This composite lattice has been revealed by means of vortex imaging techniques~\cite{Bolle1991,Grigorieva1995,Vlasko-Vlasov2002,Tokunaga2003}. Two relevant parameters for
the physics of the vortex lattice in these compounds are $\gamma$,
the electronic anisotropy, and $s$, the separation between the
CuO$_2$ planes. The coupling between pancake vortices of
adjacent planes is produced by electromagnetic and Josephson
interactions, the latter being dominant when the Josephson length
$\lambda_{\rm J}=\gamma s$ is smaller than the superconducting penetration depth $\lambda$. Layered cuprates present a different balance between these two interactions strongly depending  on the oxygen-doping. ~\cite{Vlasko-Vlasov_2015,Correa}

This composite nature  makes reaching a unified description of vortex matter in
layered high-$T_{\rm c}$ superconductors under tilted fields rather challenging.  This basic issue is of crucial
relevance for finding paths for enhancing pinning and critical currents. The latter is highly desirable for technological applications, and a good strategy to
achieve these conditions is the generation of effective pinning
landscapes. To this end, introducing columnar defects (CDs) produced by heavy-ion
irradiation is one of the most  effective approaches~\cite{Civale1991,Thompson1992,Thompson1997,Civale1999,Taphou2022,Ammor2004,Kwok}.

The case of parallel CD running along the $c$-axis (perpendicular to the CuO$_2$ planes) has been widely studied in the extremely layered Bi$_2$Sr$_2$CaCu$_2$O$_{8+\delta}$ compound that we study here. Their effect
on the structural properties of the vortex lattice as well as on the
first-order vortex melting transition is vastly reported in the literature.~\cite{Verdene,Tonomura2001,Dai1994,CejasBolecek2016} These works have settled the general consensus that this type of defect
is a very effective pinning center when the field is applied
parallel to them.  CDs are even more
efficient pinning centers when the field is applied in the $c$-axis but defects are tilted with
respect to this direction since the pinning energy is proportional to the cross-sectional area of
the defects in the CuO$_2$ planes~\cite{Ammor2004,Hebert1998,Ishikawa2004}.
Moreover, two sets of crossed  CDs
tilted with respect to the $c$-axis are more efficient to improve
the critical current than a parallel set of
defects~\cite{Schuster1996,Hebert2000}. Thus, the
spatial distribution and orientation of CD with respect to the
CuO$_2$ planes are key for their pinning efficiency~\cite{Kato2008}.

In extremely layered cuprates, when the field is
applied tilted from the $c$-axis, the pinning of pancake vortices
results from the combination of the pinning exerted by CD and by the
CuO$_2$ planes. Then, the  vortex arrangement that minimizes
the free energy in different field and temperature ranges is not
necessarily that of pancakes located inside CD. In some cases, a
staircase vortex configuration~\cite{Silhanek1999} can be energetically favorable. It consists of parts of the flux line trapped in different defects, linked by unpinned or weakly pinned segments.
 In addition, for tilted fields, pancake vortices
produce a pinning effect that indirectly affects the dynamics of
Josephson vortices~\cite{Koshelev2006}.

In order to probe the efficiency of pinning by CD for different field directions in extremely layered cuprates, here, we study the angular dependence of the ac screening of vortex matter
in Bi$_2$Sr$_2$CaCu$_2$O$_{8+\delta}$ samples with various
densities of  CDs tilted 45$^{\circ}$
from the $c$-axis.   We apply local ac Hall magnetometry in order to obtain information on the sustainable current of the vortex system, a measure of the critical current taking into account creep effects~\cite{vanderBeek1995a}. In particular, we measure the magnetic transmittivity $T'$  that reflects the sustainable current flowing in the sample for a given ac field $J = (1/\pi) \arccos(2T' -1)$~\cite{vanderBeek1995a}.
 The angular variation of ac screening has been
previously used to reveal the localization of vortices on CDs aligned along the sample
$c-$\,axis~\cite{vanderBeek2001}. The main result of this work is that
transmittivity curves are asymmetric and present dips revealing angularly selective enhanced screening. A dip in $T'$ indicates a maximum in the sustainable current due to increased pinning in the particular angular location. The enhancement of screening is registered close to the direction of CDs, but extra dips at smaller angles are also detected.  The detection
of a maximum screening  in an angular direction not aligned
with CDs is unexpected and suggests the existence of staircase pancake vortex configurations that are effectively trapped by tilted CDs.

%%%%%%%%%%%%%%%%%%%%%%%%%%%%%%%%%%%%%%%%%%
\section{Materials and Methods}

Columnar defects were produced in platelet-like optimally doped
Bi$_2$Sr$_2$CaCu$_2$O$_{8+\delta}$ single crystals grown by the
traveling-solvent floating-zone technique~\cite{Li1994}.
Sub-millimeter size and $\sim$20\,\textmu m thick samples were
irradiated with 6\,GeV Pb ions at GANIL, France. The direction of
the  beam of ions was rotated from the sample $c-$\,axis in order to
create columnar tracks of defects tilted 45$^{\circ}$ from the $c$-axis. The
penetration range of the 6\,GeV Pb ions largely exceeds the  sample
thickness divided by the sinus of the tilt angle.   The flux of irradiation was chosen to
generate various densities of defects corresponding to matching
fields when the field is applied in the direction of CDs of $B_{\Phi}$
= 50, 200 and $2000$\,G , with $B_{\Phi}$ as the field value at which
the density of CD equals that of the vortices. When the field is applied
in the $c$-axis direction, the density of the vortices equals that of CDs
in a plane parallel to the $ab$-plane for a smaller density of
$B^{c}_{\Phi}=B_{\Phi}/\sqrt{2} = 35.5, 141.9$ and $1419$\,G. These
densities of CD are such that the separation between defects is
smaller ($B_{\Phi}=2000$\,G) and larger ($B_{\Phi}=200$, 50\,G) than $\lambda_{\rm J} = \gamma s$$\sim$0.25\,\textmu m for Bi$_2$Sr$_2$CaCu$_2$O$_{8+\delta}$ with an
anisotropy $\gamma$$\sim$160 and a spacing between CuO$_2$ planes of
$s=15$\,\AA.

\begin{figure}[ttt]
    \includegraphics[width=\columnwidth]{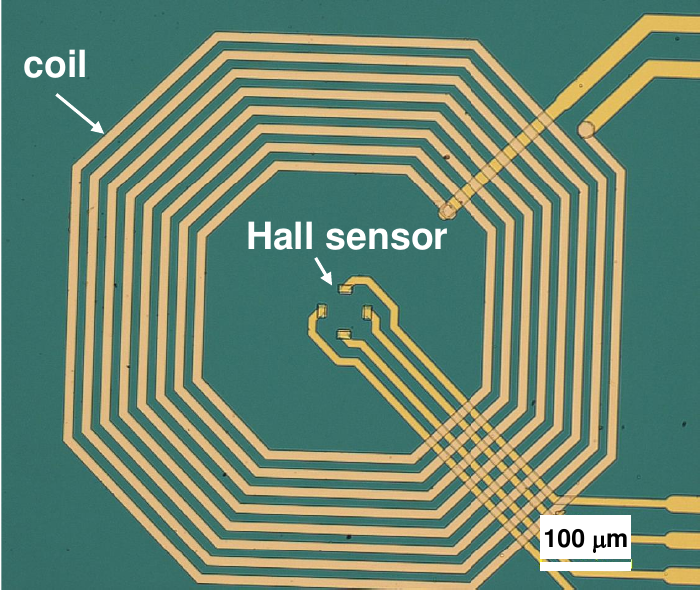}
    \caption{Hall sensor setup used in the local transmittivity measurements of this work. Detail of the
     $10\times10$\,\textmu m$^2$ working-area  Hall sensor surrounded by an on-chip
      coil that generates the ripple field for ac measurements. The
      platelet-like superconducting sample (not shown)  is glued with apiezon-N grease parallel to the chip and on top of the Hall
      probe and the coil. The sensor detects the sample stray field, $B^{\perp}$, pointing in the direction perpendicular
       to its $ab$-plane.}
    \label{fig:Sonda}
\end{figure}

The sample is mounted on top of a chip containing a Hall sensor plus a planar coil. The
sample is located with its $ab$-plane roughly parallel to the sensor, glued with Apiezon-N grease to
improve thermal contact. The component of the local stray field at the sample surface pointing perpendicular to the
$ab$-plane,
$B^{\perp}$, is detected by a   Hall sensor with a $9 \times 9$\,\textmu m$^2$ working area photolithographically fabricated  from  a so-called AlGaAs/InGaAs/GaAs pseudomorphic heterostructure with a reduced 2D electron gas density in order to achieve a higher sensitivity amounting to $\sim$146\,m$\Omega$/G.
In order to perform ac magnetometry measurements, the sensor is surrounded by an on-chip coil generating the ripple $h_{\rm ac}$ field.
This coil is made up of 8-turn 0.6\,\textmu m thin Au film deposited over a silicon oxinitride dielectric layer, with a radius ranging from 160 to 300\,\textmu m µm and a pitch of 20\,\textmu m.
Figure\,\ref{fig:Sonda} shows a
picture of a typical  chip containing the Hall sensor and the coil.
This chip is then attached into a sample holder that is thermally
coupled to the thermometer registering the temperature.

\begin{figure}[ttt]
\includegraphics[width=0.9\columnwidth]{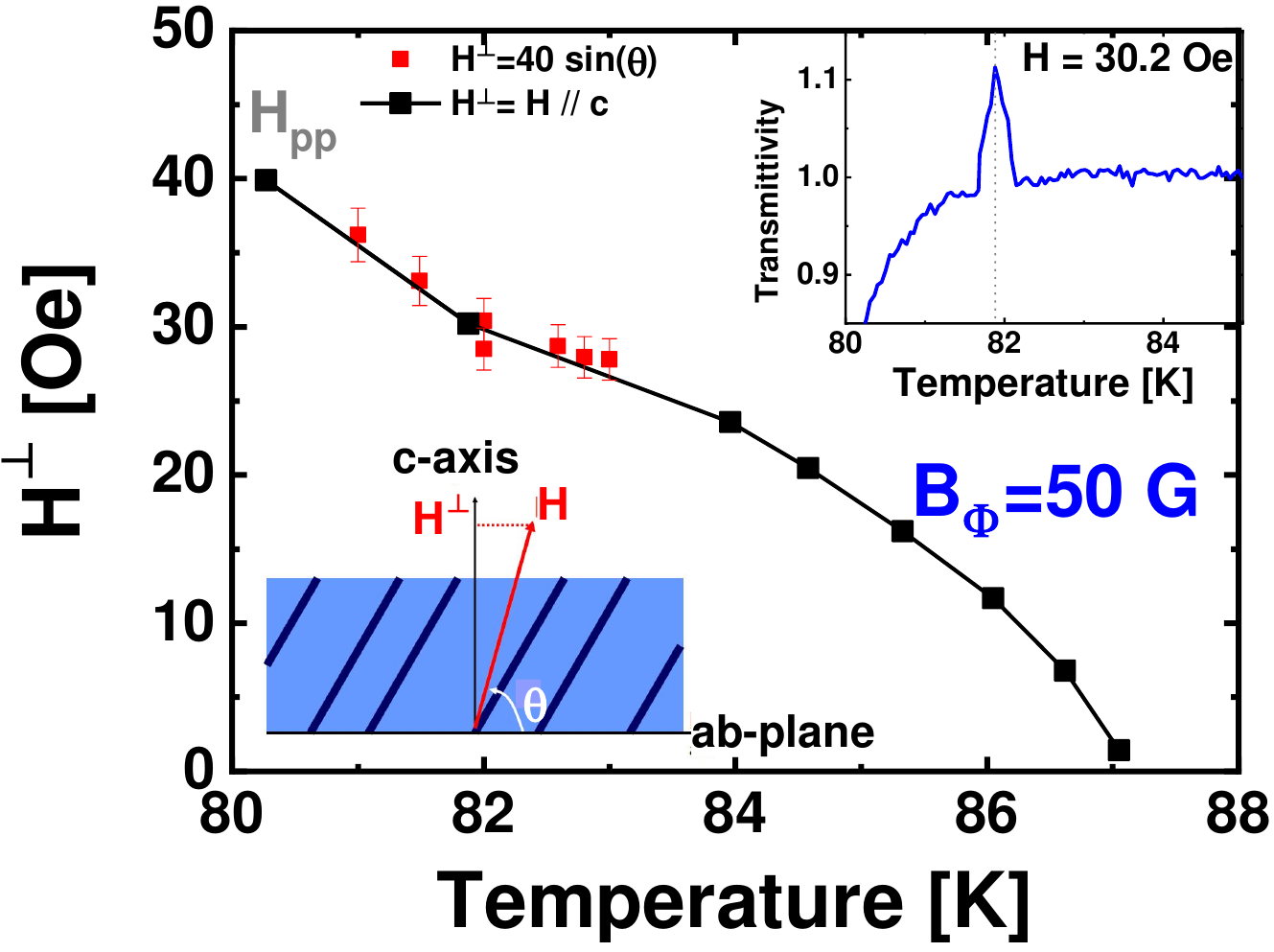}
\caption{Phase diagram and angular dependence of transmittivity $T'$ for
a Bi$_2$Sr$_2$CaCu$_2$O$_{8+\delta}$ sample with columnar
defects with a matching field of $B_{ \Phi}=50$\,Oe tilted 45$^{\circ}$ from the $c$-axis. First-order melting
line $H_{\rm pp}$ obtained from the location of the paramagnetic
peak detected in $T'$ versus temperature measurements at fixed field (see top insert for example). Black dots correspond to measurements
performed with the field applied in the $c$-axis direction whereas
red dots correspond to the $c$-axis component $H^{\perp}$ of the
field applied at an angle $\theta$ from the $ab$-plane in angular
measurements. Top insert: $T'$ data
obtained in the proximity of the paramagnetic peak (dashed line) for
30.2\,Oe applied in the $c$-axis. Bottom
insert: Schematics of the tilted columnar defects (thick black
lines) generated  at an angle of 45$^{\circ}$ with respect to the
$c$-axis and the $ab$-plane in a plane running along the long side of the
sample. The dc field $H$ is applied in the
plane of the defects at an angle $\theta$ with respect to the
$ab$-plane of the sample.} \label{fig:DiagF}
\end{figure}

Local magnetic measurements are performed  by applying a dc magnetic
field $H$ at an angle $\theta$ from
the $ab$-plane of the samples and with a component $H^{\perp}$ along
the $c$-axis; see the insert to Figure\,\ref{fig:DiagF}.  The ac
transmittivity measurements are performed, applying, in addition, a
ripple field $h_{\rm ac}$ that is parallel to the $c$-axis.
 Angular measurements are performed by
sweeping the external magnetic field within the plane defined by the $c$-axis and the direction of CDs; see the bottom-left insert to Figure\,\ref{fig:DiagF}.  As a consequence, the direction of $H$ varies, generating
  a simultaneous    sweep of the $c-$axis, $H^{\perp}$,  and of
  the $ab$-plane components. We perform these measurements in a controlled way using a step motor, allowing us to reach
half a degree resolution. Positive $\theta$ corresponds to angles
going from the $ab$-plane towards the $c$-axis. All measurements
presented here start with the field applied at $\theta =
-90$$^{\circ}$ and then the field is rotated counterclockwise
towards the $ab-$ plane with $\theta=0$ and finally towards
$\theta=90$$^{\circ}$.
$h_{\rm ac}$ has a magnitude ranging  0.5--1\,Oe rms and frequencies
between 0.7 and 1000\,Hz.

Hall magnetometry allows the detection of $B^{\perp}$, the local
magnetic induction of the sample in the direction perpendicular to
the sensors. In ac measurements, we detect the first harmonic of
$B^{\perp}$ by means of a digital signal-processing lock-in
technique.    The transmittivity $T'$ is obtained by  normalizing
the in-phase component of the first-harmonic signal $B'$ such that
$T'(T,\theta)=[B'(T,\theta) - B'(T \ll T_{\rm
c},\theta)]/[B'(T>T_{\rm c},\theta) - B'(T \ll T_{\rm
c},\theta)]$~\cite{Dolz2014}. Then, $T'=1$ in the normal phase
     for $T>T_{\rm c}$ and $T'=0$ when full screening takes place at low
     temperatures well within the superconducting phase.
In order to normalize the angular measurements, we first record
$B'(\theta)$ for a given applied $H$ and for temperatures $T
> T_{\rm c}$ and $T << T_{\rm c}$. We then use these normalization
curves to obtain      $T'(T,\theta)$ at the same magnitude of $H$
that rotates in the plane of the CDs. Once a value of  $B'$ at a given
direction of the field is measured, the magnet is rotated by
0.5$^{\circ}$, and the next point of the $B'(\theta)$ curve is
measured. Afterwards, by applying the mentioned normalization, we obtain
the $T'(\theta, H, T)$ curves. The transmittivity is  a magnitude
extremely sensitive to discontinuities in the local induction
associated with first-order magnetic transitions such as, for instance, the
vortex melting transition~\cite{Dolz2014,Dolz2015b}. In the $T'$
angular measurements that we present here, we introduce an ac ripple
field parallel to the $c$-axis on the background of an oblique dc
field; namely, we measure the in-plane screening current affected by
an out-of-plane dc field~\cite{Dolz2014}.

\begin{figure}[ttt]
\includegraphics[width=0.7\columnwidth]{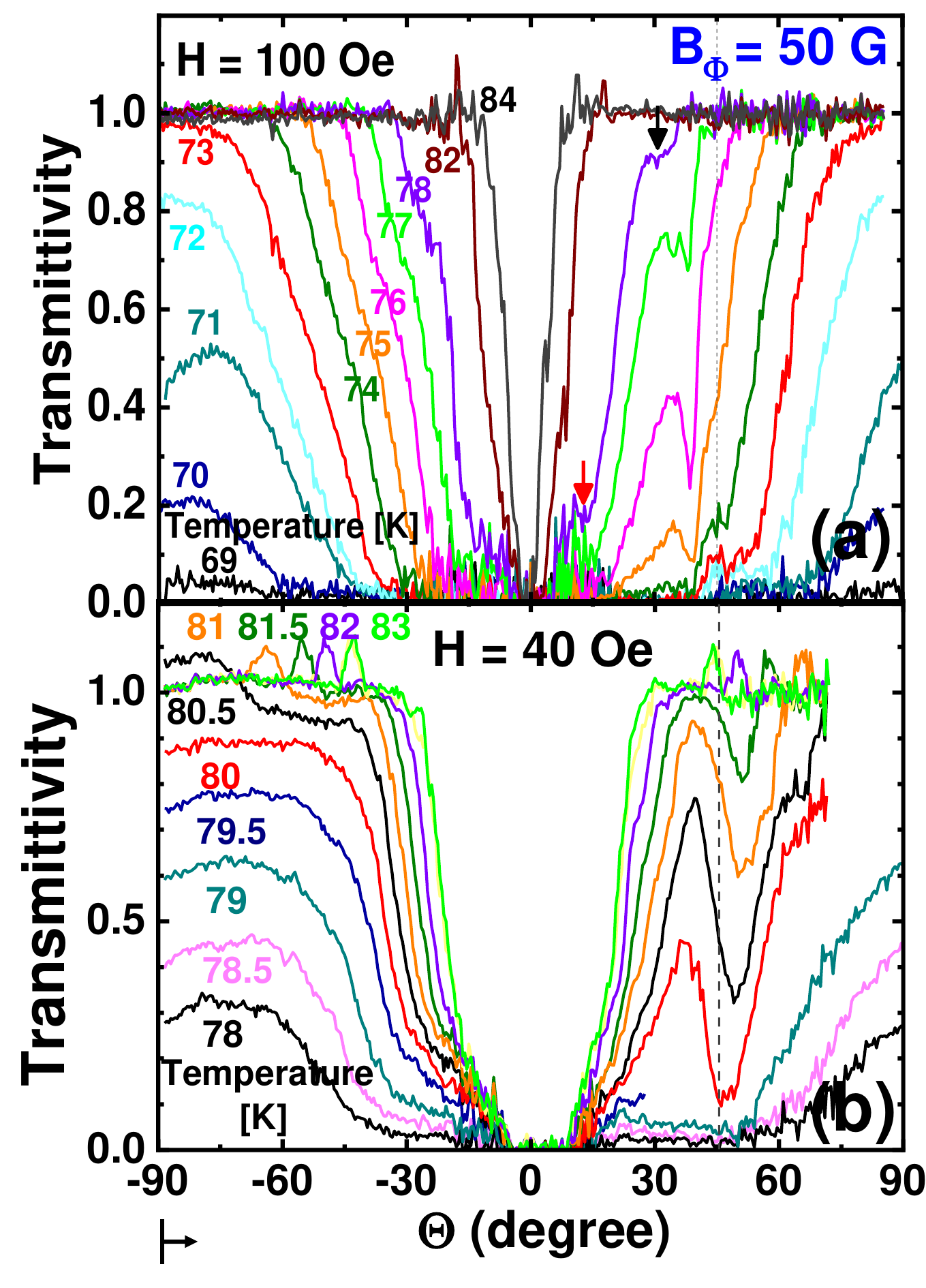}
\caption{Angular dependence of ac transmittivity $T'$ for a
Bi$_2$Sr$_2$CaCu$_2$O$_{8+\delta}$
sample with tilted columnar
defects (at 45$^{\circ}$ from the $c$-axis) and a density
corresponding to $B_{\Phi}=50$\,G. $T'$ at various
temperatures close to the melting line for applied fields of (\textbf{a})
$100$\,Oe and (\textbf{b}) $40$\,Oe. Positive (negative) angles
correspond to the anti$-$clockwise (clockwise) direction from the
$ab$-plane. The applied field is swept from the $c$$-$axis direction,
$\theta=90$$^{\circ}$, towards the $ab$$-$plane direction,
$\theta=0$$^{\circ}$, and beyond (see black arrow at the bottom left). Dashed lines indicate the
direction of defects. Measurement temperatures are indicated using the same
color code as the curves. The black arrow indicates the dip detected close to the direction of the columnar defects whereas the red one indicates a dip registered at a smaller angle.} \label{fig:Fig3}
\end{figure}
%%%%%%%%%%%%%%%%%%%%%%%%%%%%%%%%%%%%%%%%%%
\section{Results}

\begin{figure}[ttt]
\includegraphics[width=0.7\columnwidth]{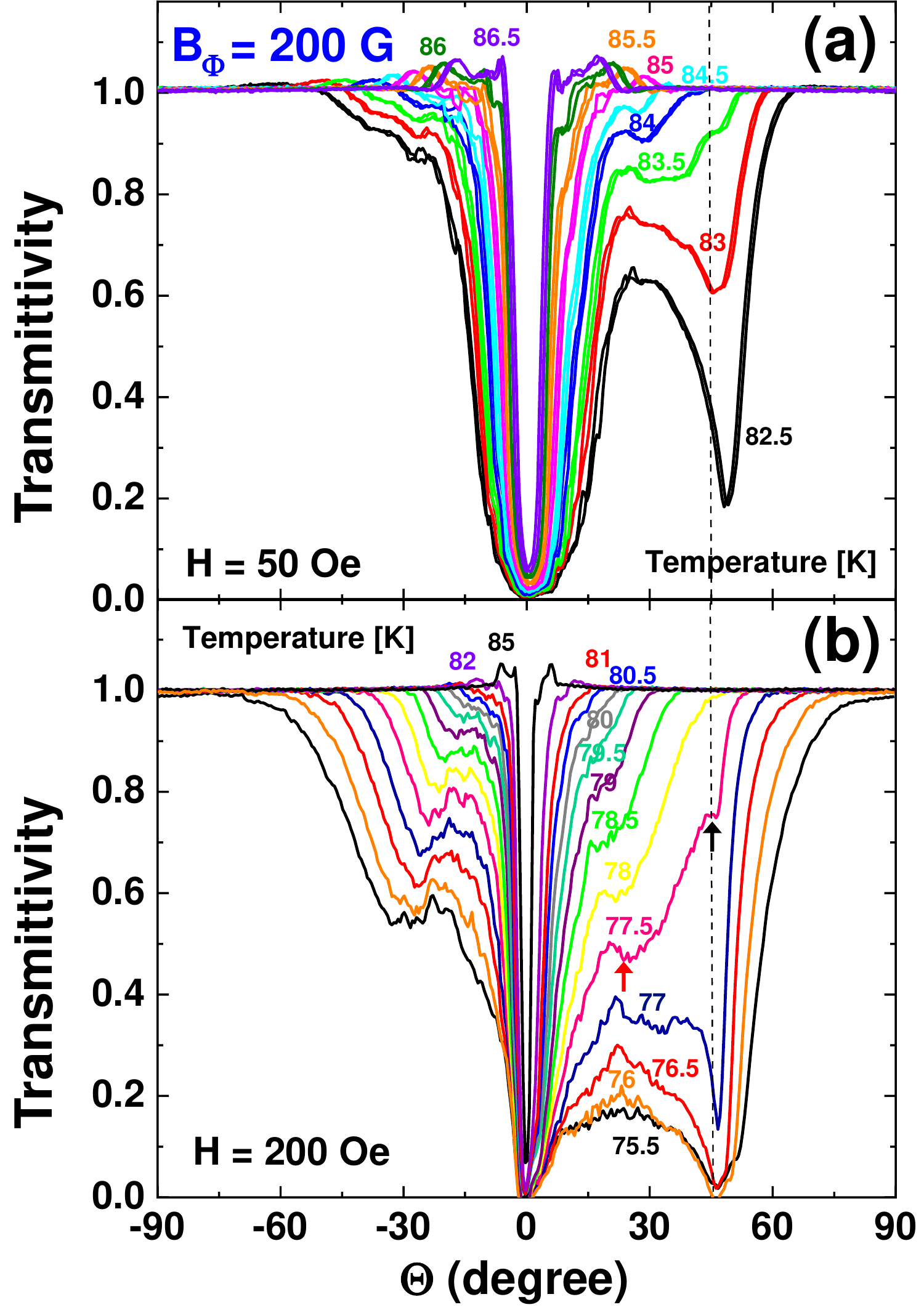}
\caption{Angular dependence of ac transmittivity $T'$ for a
Bi$_2$Sr$_2$CaCu$_2$O$_{8+\delta}$ sample with tilted columnar
defects (at 45$^{\circ}$ from the $c$$-$axis) and a density
corresponding to $B_{\Phi}=200$\,G. $T'$ at various temperatures  for applied fields of (\textbf{a}) $50$ and (\textbf{b}) $200$\,Oe. Positive (negative) angles correspond to the
anti$-$clockwise (clockwise) direction from the $ab$-plane. The
applied field is swept from the $c$$-$axis direction,
$\theta=90$$^{\circ}$, towards the $ab$$-$plane direction,
$\theta=0$$^{\circ}$, and beyond. The dashed lines indicate the
direction of the columnar defects. Labels corresponding to the temperatures  are indicated with the same
color code as the curves. The black arrow indicates the dip detected close to the direction of the columnar defects whereas the red one indicates a dip registered at a smaller angle.} \label{fig:Fig6}
\end{figure}

\begin{figure}[ttt]
\includegraphics[width=0.73\columnwidth]{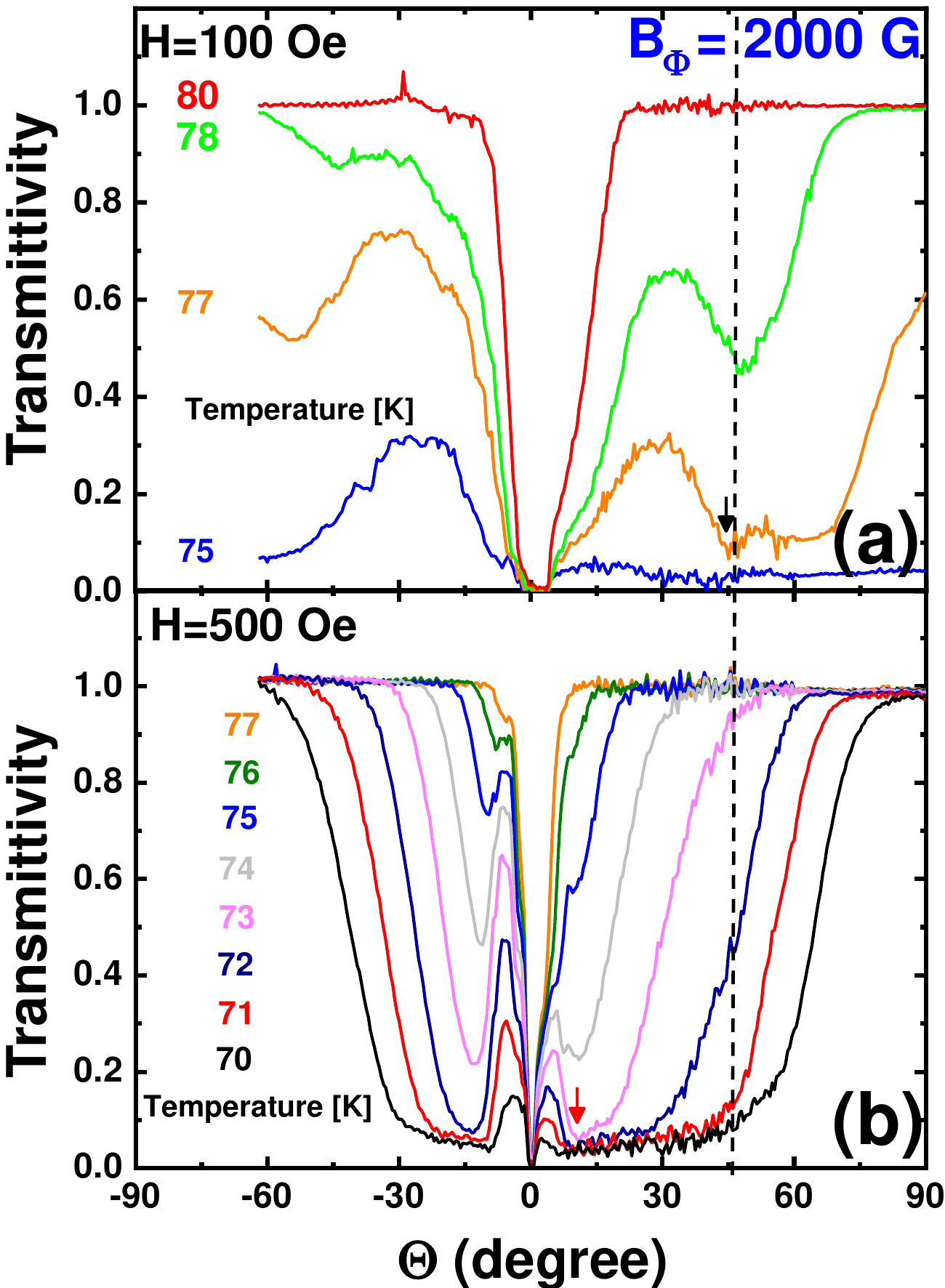}
\caption{Angular  dependence of ac transmittivity $T'$  for a
Bi$_2$Sr$_2$CaCu$_2$O$_{8+\delta}$ sample with tilted columnar
defects (at 45$^{\circ}$ from the $c$$-$axis) and a density
corresponding to $B_{\Phi}=2000$\,G. $T'$ at
different temperatures  for applied fields of (\textbf{a}) $100$ and (\textbf{b}) 500\,Oe. Positive (negative) angles correspond to the
anti$-$clockwise (clockwise) direction from the $ab$$-$plane. The
applied field is swept from the $c$$-$axis direction,
$\theta=90$$^{\circ}$, towards the $ab$$-$plane direction,
$\theta=0$$^{\circ}$, and beyond. The dashed lines indicate the
direction of the columnar defects. Labels corresponding to the measurement temperatures are indicated with the same
color code as the curves. The black arrow indicates the dip detected close to the direction of the columnar defects whereas the red one indicates a dip registered at a smaller angle.} \label{fig:Fig5}
\end{figure}

\begin{figure}[ttt]
\includegraphics[width=0.65\columnwidth]{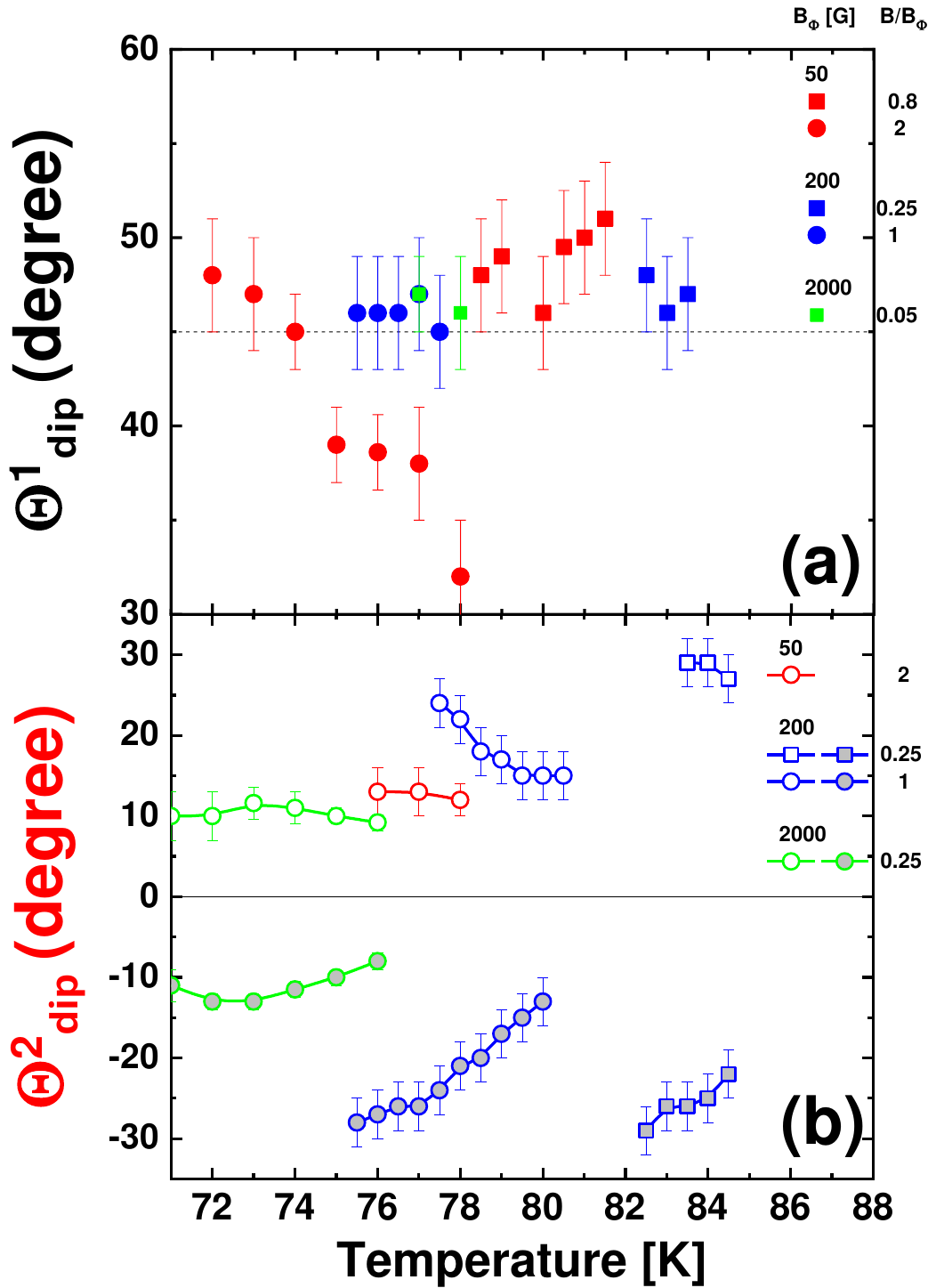}
\caption{Angular location of the dips in transmittivity for
Bi$_2$Sr$_2$CaCu$_2$O$_{8+\delta}$ samples with tilted columnar
defects (at 45$^{\circ}$ from the $c$-axis) with densities
corresponding to $B_{\Phi}=50$ (red points), $500$ (blue points) and
$2000$\,G (green points). Data are obtained from the curves of
Figures \ref{fig:Fig3}  to \ref{fig:Fig5}  at various temperatures and
different applied fields. The label indicates the ratio between the
field applied to measure and the matching field, $B/B_{\Phi}$,
namely, the vortex to CD number ratio. (\textbf{a}) Angular location of the dip in transmittivity detected close to the columnar defects direction, $\Theta^{1}_{\rm dip}$. (\textbf{b}) Angular location of the sharper dips located at smaller angles $\Theta^{2}_{\rm dip}$ roughly symmetric with respect to the $ab$$-$plane.} \label{fig:Fig7}
\end{figure}

 The transmittivity
versus temperature measurements performed with the field applied
parallel to the $c$-axis reveal the presence of paramagnetic peaks;
see the example of the top insert to Figure\,\ref{fig:DiagF}. The
temperature--location of these peaks is
independent of the frequency of the ripple field used to measure. In
pristine samples, paramagnetic peaks with such phenomenology are
considered indicative of the first-order melting transition since
a peak in $T'$ comes from a peak in $B'$, and then is the consequence
of a jump in $B$~\cite{Dolz2014,AngularDepFOT}. The
location of these peaks in the field-temperature phase diagram is
indicated with black dots in Figure\,\ref{fig:DiagF} generating the $H_{\rm
pp}$ line that indicates the location of the  first-order solid-to-liquid vortex transition.  The paramagnetic peaks in the $B_{\Phi} = 50$\,G sample are clearly detected
down to 81\,K and up to 40\,Oe; for larger fields or smaller
temperatures, we are not able to detect them due to  the sudden decrease in
$T'$ associated with the screening in the vortex solid phase. These
paramagnetic peaks are also resolved at high temperatures in the sample with
$B_{\Phi}=200$\,G.

Figure\,\ref{fig:Fig3} shows angular-dependent transmittivity data,
 namely, $T'$ vs. $\theta$ for the $B_{\Phi} =50$\,G sample at applied fields $H$ of (a) 100 and (b) 40\,Oe, and a  ripple field of  0.7\,Oe  and 11\,Hz. In this set of measurements, the paramagnetic
   peaks are clearly  detected in the $H=40$\,Oe case and appear more noisy in the $H=100$\,Oe case.
   Figure\,\ref{fig:Fig3}b shows the succession of
  paramagnetic peaks close to the $T'=1$ value: they are located
  at smaller $\theta$, namely, smaller $H^{\perp}$, for larger temperatures.
  The peaks are roughly symmetrically detected at a given angle $\theta$
   with respect to the $ab$-plane.
  Considering the value of the component $H^{\perp}=H \sin{\theta}$ and the
  temperature at which these peaks are detected, we obtain the red dots shown
  in Figure\,\ref{fig:DiagF} for $H=40$\,Oe. These red points
coincide with the $H_{\rm pp}$ black points.
These  results suggest that for a low density of CD, the relevant variable for the occurrence  of the first-order transition in tilted fields for extremely anisotropic superconductors is the pancake vortex density, i.e., the $c$-axis component of the applied field.  Thus, for a moderate CD
   density like $B_{\Phi}=50$\,G, the nature of the first-order transition line detected at high temperatures
   in Bi$_2$Sr$_2$CaCu$_2$O$_{8+\delta}$ is not significantly altered with respect
   to the case of pristine samples. Indeed, note that when comparing our results with those of Ref.\,\cite{Composite_PV-JV}, it is clear that the in-plane component is not large enough to produce a significant displacement of the field at which the transition occurs.

In the studied temperature range, the  transmittivity vs. 
$\theta$ curves are asymmetric with respect to both the $ab$-plane
and the CD direction. This asymmetry is partly due to a ubiquitous feature in $T'$: a dip in the angles close but not exactly equal to the direction of CDs, visible only at positive angles. These dips are indicated with black arrows in Figures\,\ref{fig:Fig3}--\ref{fig:Fig5} for the three studied CD densities and appear at angles $\Theta^{1}_{\rm dip}$. Local dips in ac transmittivity are associated with an enhancement of screening.
The dips are observed at temperatures larger than 70\,K, in a  range of
fields  below and above $B_{\Phi}$.
The angles $\Theta^{1}_{\rm dip}$ are larger than the
ones at which we detect the paramagnetic peaks in the samples with $B_{\Phi}=50$ and 200\,G
associated to the first-order
transition. Then, these dips occur for
components $H^{\perp}$ smaller than
$H_{\rm pp}$; namely, this enhanced screening
takes place within the solid vortex phase.

Figure\,\ref{fig:Fig7} (a) summarizes the location of these dips for all the studied samples at different fields and temperatures.
For vortex densities $B/B_{\Phi} \leq1$, this enhancement of screening is detected close to, but not exactly aligned with, the direction of CD, namely, $\Theta^{1}_{\rm dip}$$\sim$$45^{\circ}$. Then, this feature is quite possibly a manifestation of enhanced screening due to CD strongly pinning pancake vortices when the field is aligned close to the direction of defects.   In the case when $B/B_{\Phi} =2$, for temperatures greater than 75\,K, $\Theta^{1}_{\rm dip}$ significantly decreases upon warming from the low-temperature $\sim$$45^{\circ}$ value.

The transmittivity curves for the three studied CD densities also present extra dips in $T'$ that are generally less pronounced than the ones located at $\Theta^{1}_{\rm dip}$. These dips,  indicated with red arrows in Figures \,\ref{fig:Fig3}, \,\ref{fig:Fig6}, and \ref{fig:Fig5}, are detected at angles $|\Theta^{2}_{\rm dip}|< 45^{\circ}$ for $B/B_{\Phi}$ larger and greater than one. These dips are observed for positive as well as negative angles,  appearing quite symmetric with respect to the $\theta = 0$ direction; see Figure\,\ref{fig:Fig7}b for further details. In the case of the  $B_{ \Phi}=50$\,G sample, the negative $\Theta^{2}_{\rm dip}$ values are not reported since the transmittivity curves are too noisy to provide a good estimation of this magnitude, though the dips at negative angles are faintly observed in the curves.  Thus, the dips at $\Theta^{2}_{\rm dip}$  are a manifestation of an enhanced screening that occurs roughly symmetrically with respect to the $ab$-plane. In addition, $|\Theta^{2}_{\rm dip}|$ decreases upon increasing the temperature, concomitant with the dips becoming less pronounced upon warming.

%%%%%%%%%%%%%%%%%%%%%%%%%%%%%%%%%%%%%%%%%%
\section{Discussion}

Tilted fields in extremely layered superconductors can give rise to a composite lattice phase   where pancake vortices coexist with Josephson vortices, or to tilted chains of pancake vortices~\cite{Composite_PV-JV,Koshelev_2005}. A transition between both types of arrangements is reported by means of ac transmittivity measurements~\cite{Composite_PV-JV}. In our experimental data on the angular dependence of transmittivity, we explore a continuous set of ratios between in-plane and $c$-axis field components,  $H_{||}/H_{\perp}$. The dips in $T'$ are detected within a $H_{||}/H_{\perp}$ range that does not coincide with the composite-to-tilted lattice transition experimentally reported in pristine Bi$_2$Sr$_2$CaCu$_2$O$_{8+\delta}$~\cite{Composite_PV-JV}. Therefore, the dips  at $\Theta^{1}_{\rm dip}$ and $\Theta^{2}_{\rm dip}$   are not associated to this structural transition in vortex matter.

Another phenomenon relevant to tilted fields in layered superconductors is the so-called \textit{lock-in} angle, which refers to the alignment of the vortex lattice with the $ab$-planes, creating a purely Josephson-vortex array~\cite{Blatter,Lockin}. Considering the typical parameters for Bi$_2$Sr$_2$CaCu$_2$O$_{8+\delta}$, the lock-in angle where this arrangement is favorable should lie below 2$^{\circ}$. This angle is much smaller than the ones at which the dips in $T'$ are detected, and thus they cannot be associated to this phenomena either.

Therefore, the dips detected in our experimental range at $\Theta^{1}_{\rm dip}$ and $\Theta^{2}_{\rm dip}$ are more likely associated to a change in configuration between different arrangements within the composite lattice phase. The dips in $T'$ are then the fingerprint of enhanced screening in a selected angular range associated to an improved pinning efficiency induced by the CD arrangement  in the composite lattice phase. Since for $B/B_{\Phi} \leq 1$ the magnitude of $\Theta^{1}_{\rm dip}$$\sim$45$^{\circ}$, these dips are clearly related to pinning by the CD of a vortex arrangement made of tilted stacks of pancake vortices following the direction of CD that might eventually have a moderate deformation induced by the in-plane Josephson vortices. The fact that $\Theta^{1}_{\rm dip}$ is not exactly $45$$^{\circ}$ degrees indicates that once this configuration is reached at the vicinity of the CD direction upon increasing $\theta$, the system remains efficiently pinned by the defects even though the angle of the applied field continues increasing.

As mentioned, when $B/B_{\Phi} >1$, $\Theta^{1}_{\rm dip}$ significantly decreases upon warming from the low-temperature $\sim$$45^{\circ}$ value. This is an indication that for a larger density of vortices than that of CD,  the system arranges in a configuration where only a fraction of the pancakes of individual vortices are pinned by CD and the rest lie outside of them.  Then the fact that $\Theta^{1}_{\rm dip}$ decreases upon warming implies that the enhanced-screening configuration of the system
occurs for a larger $H_{||}$ and then an increasing number of Josephson vortices.
This is suggestive of not only CD but also Josephson vortices having a  relevant role in determining the enhanced-screening configuration of the system and deserves further investigation.

As for the dips detected at $\Theta^{2}_{\rm dip}$, since this angle is rather small, they seem to correspond to a staircase vortex configuration, where only a fraction of pancakes of individual vortices is tilted along the direction of CD. The more remarkable property of this feature is that the dips appear in pairs located almost symmetric with respect to the $ab$-plane direction. This is an indication that the enhanced screening in this angular range might be due to a combination of the pinning of CD with the pinning exerted by Josephson vortices that is symmetric with respect to the $ab$-plane. Another salient property of these dips is that the modulus of $\Theta^{2}_{\rm dip}$ decreases upon warming. This finding  indicates, as in the case of $\Theta^{1}_{\rm dip}$, that
the enhanced-screening configuration of the system
occurs for a larger number of Josephson vortices upon warming.

%%%%%%%%%%%%%%%%%%%%%%%%%%%%%%%%%%%%%%%%%%
\section{Conclusions}

In conclusion, our angular transmittivity data in extremely layered
Bi$_2$Sr$_2$CaCu$_2$O$_{8+\delta}$ samples with tilted CD indicate
that the screening of a ripple field is enhanced  when the vortex
direction is around that of CDs but an extra enhancement of screening is detected at small angles, almost symmetric with the $ab$-plane direction.  This particular angularly selective enhanced screening has not been reported in other superconductors and arises due to the efficiency of  CDs in pinning a composite lattice of vortices nucleated in an extremely layered material in tilted fields. For directions close to the CD, as expected, pancake vortices profit at maximum pinning, being located at the defects. Strikingly,  screening is also enhanced at smaller angles where the composite lattice is efficiently pinned, probably by adjusting itself into a staircase configuration to take better advantage of the pinning landscape.

%%%%%%%%%%%%%%%%%%%%%%%%%%%%%%%%%%%%%%%%%%

%%%%%%%%%%%%%%%%%%%%%%%%%%%%%%%%%%%%%%%%%%

This research was funded by  the National Council of Scientific
and Technical Research of Argentina (CONICET)
through grant PIP 2021-1848, by the Argentinean
Agency for the Promotion of Science and Technology
(ANPCyT) through grant PICT 2017-2182, and by the Universidad Nacional de Cuyo
research Grant 06/C008-T1.

%%%%%%%%%%%%%%%%%%%%%%%%%%%%%%%%%%%%%%%%%%

%%%%%%%%%%%%%%%%%%%%%%%%%%%%%%%%%%%%%%%%%%
%%
%%%%%%%%%%%%%%%%%%%%%%%%%%%%%%%%%%%%%%%%%%

\end{document}